\documentclass{article}
\usepackage{icrctc07}

\title{Neutrino Triggered Target of Opportunity (NToO) test run with AMANDA-II and MAGIC}
\shorttitle{Neutrino Target of Opportunity}

\authors{M. Ackermann$^1,5$, E. Bernardini$^1$, N. Galante$^2$, F. Goebel$^2$, M. Hayashida$^2$, K. Satalecka$^1$, M. Tluczykont$^1$, R.~M. Wagner$^2$, for the IceCube$^3$ and MAGIC collaborations$^4$}
\shortauthors{Ackermann and et al.}
\afiliations{$^1$DESY, Platanenallee 6, 15738 Zeuthen\\ 
             $^2$MPPMU, F\"ohringer Ring 6, 80805 M\"unchen\\
             $^3$See special section of these proceedings\\
             $^4${\tt http://magic.mppmu.mpg.de/collaboration/members}\\
             $^5$Now at SLAC, Stanford University, USA}
\email{elisa.bernardini@desy.de} 

\abstract{Kilometer scale neutrino telescopes are now being
constructed (IceCube) and designed (KM3NeT). While no neutrino flux of
cosmic origin has been discovered so far, the first weak signals are
expected to be discerned in the next few years. Multi-messenger 
(observations combining different kinds of emission) 
{investigations can enhance the discovery chance for neutrinos in case
of correlations}. One possible application is the search for time
correlations of high energy neutrinos and established signals. We show
the first adaptation of a Target of Opportunity strategy to collect
simultaneous data of high energy neutrinos and gamma-rays. 
Neutrino events with coordinates close to preselected candidate sources 
are used to alert gamma-ray observations. The detection
of a positive coincidence can enhance the neutrino discovery
chance. More generally, this scheme of operation can increase the
availability of simultaneous observations. If cosmic neutrino signals
can be established, the combined observations will allow time
correlation studies and therefore constraints on the source
modeling. A first technical implementation of this scheme involving
AMANDA-II and MAGIC has been realized for few pre-selected sources in a
short test run (Sept. to Dec. 2006), showing the feasability of the concept.
Results from this test run are shown.
}

\begin{document}

\newcommand{\Nobs}{$n_\mathrm{obs}\,$}
\newcommand{\Nbck}{$n_\mathrm{bck}\,$}
\newcommand{\Ncoinc}{$n_{\gamma}\,$}
\newcommand{\pgam}{${p_{\gamma}}\,$}
\newcommand{\tevflux}{$\mathrm{F_{VHE}}$}
\newcommand{\asmflux}{$\mathrm{F_{ASM}}$}
\newcommand{\rhs}{${R_{HS}}$}
\newcommand{\thigh}{$\mathrm{T_{high}}$}
\newcommand{\tlow}{$\mathrm{T_{low}}$}

\newcommand{\signeu}{P$_\nu$}
\newcommand{\sigtoo}{S$_{ToO}$}

\maketitle

\section{Introduction}
\label{intro}
The major aim of neutrino astrophysics is to contribute to the understanding 
of the origin of high energy cosmic rays. 
A point-like neutrino signal of cosmic origin would be an unambiguous signature 
of hadronic processes, unlike $\gamma$-rays which can also be created in
leptonic processes.
%
Neutrino telescopes are ideal instruments to monitor the sky and 
look for the origin of cosmic rays
because they can be continuously operated. 
The detection of cosmic neutrinos is however very challenging because
of their small interaction cross-section and because of a large 
atmospheric background.
%
%
{Parallel measurements using neutrino and electro-magnetic observations
(multi-messenger)
can increase the chance} to 
discover the first signals by reducing the trial factor penalty 
arising from observation of multiple sky bins and over different time periods. 
%
In a longer term perspective, the multi-messenger approach also aims 
at providing a scheme for the phenomenological interpretation of the 
first possible detections.
The Antarctic Muon and Neutrino Detector Array (AMANDA) was built
with the aim to search for extraterrestrial high energy neutrinos 
\cite{amanda}. The Major Atmospheric Gamma Imaging Cherenkov telescope 
(MAGIC) is a current generation $\gamma$-ray telescope that
operates in the northern hemisphere at a trigger energy threshold of 60\,GeV 
\cite{magic}.

\vspace{-0.2cm}
\subsection{Neutrino Target of Opportunity test run}
\label{sec:NToO}

The neutrino target of opportunity (NToO) test run described here was 
defined as a cooperation between the AMANDA (neutrinos) and MAGIC
($\gamma$-rays) collaborations \cite{bernardini:2005a}.
Each time a neutrino event was detected from the direction 
of a predefined list of objects,
a trigger was sent to the $\gamma$-ray telescope.
MAGIC then tried to 
observe the object within a predefined time window
after the neutrino trigger. 
{The primary goal of the NToO approach is to achieve
simultaneous neutrino/$\gamma$-ray observations.
This can be realized by triggering
follow-up observations of interesting neutrino events,
such as multiplets within a short time window or 
very high energy events, therewith assuring 
$\gamma$-ray coverage for these neutrino events.
Multiplets are very seldom in AMANDA-II observations (low statistics). 
We therefore implemented a test run based on single high energy neutrino events
from pre-defined directions.} These events are most likely due 
to atmospheric neutrino background.
The test run took place between 27th of September and 27th of November 
2006 and its purpose was to test the
technical feasability of the NToO strategy. 
The AMANDA-II DAQ data at the South Pole passed through 
an online reconstruction
filter that selected up-going muon tracks and provided 
a monitoring of the data quality. 
Whenever a neutrino event was reconstructed within a few degrees of
one of the selected sources and passed the
data quality criteria, a message was sent 
via e-mail to the MAGIC shift crew. The message contained the time of the
event, the source name and the reconstructed angular distance from the source.
If possible (day/night duty cycle), the object was then observed 
with the MAGIC telescope within 24 hours {for a duration of 1 hour}.
A coincidence is counted when a $\gamma$-ray high state (flare) is measured
in these observations.
A $\gamma$-ray flare can be defined as an
observation above a predefined threshold flux ${F_{thr}}$.
The individual thresholds were chosen either based on the MAGIC
sensitivity or in case of Mrk\,421 to a conservatively low value for which
the probability to observe a high state as defined above
would be {of the order of few percent}.

\subsection{An example analysis: Blazars}
\label{analysis}
\begin{figure}[ht]
\centering
\includegraphics*[width=0.42\textwidth,angle=0,clip]{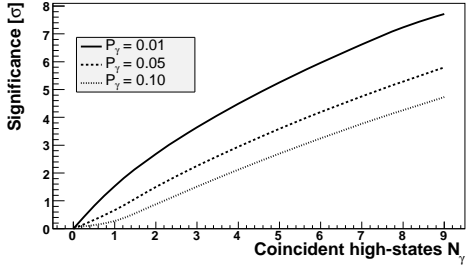}
\vspace{-0.3cm}
\caption{\label{significance}Significance of simultaneous neutrino/gamma-ray observations
         vs. the number of observed coincidences, given for
         different values of \pgam (Equation~\ref{prob}). Here, \Nobs = \Nbck = 10 was assumed.}
\end{figure}

\begin{figure}[ht]
 \centering
 \includegraphics*[width=0.42\textwidth,angle=0,clip]{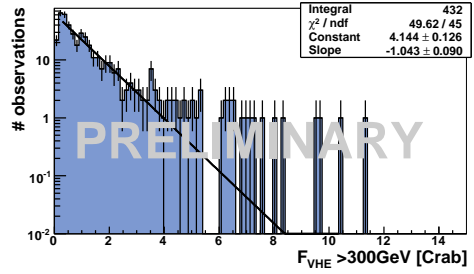}
 \vspace{-0.3cm}
 \caption{\label{fluxstates}Distribution of flux states above 300\,GeV 
          of 15 years of VHE observations of Mrk\,421 \cite{tluc}.}
\end{figure}

\begin{figure}[ht]
 \centering
 \includegraphics*[width=0.42\textwidth,angle=0,clip]{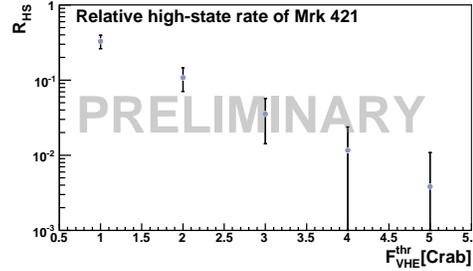}
 \vspace{-0.3cm}
 \caption{\label{hsrate}High-state rate calculated by applying equation~\ref{hsr} to
          the fit of the distribution of fluxstates in Figure~\ref{fluxstates}.}
\end{figure}

A stand-alone neutrino analysis can only yield a significant
result if an excess above the expected atmospheric background is observed. 
In the multi-messenger framework, 
the observation of a number of neutrino events in coincidence
with gamma-ray high states can be an indication
for a neutrino/$\gamma$-ray correlation.
If this correlation is incompatible with
the chance probability for coincidence
with atmospheric neutrinos
such an observation would be evidence at the same time
for a cosmic origin of the neutrino events and
a hadronic nature of the gamma-ray signal.
In this scheme for the interpretation of data a statistical 
test was defined before the measurements. 
Under the hypothesis that all the neutrinos 
detected from the direction of the source are atmospheric,
the chance probability of detecting at least \Nobs neutrinos 
and observing at least \Ncoinc coincident gamma-ray flares
is given by:
{
\begin{equation}\label{prob}
\small
P=\sum_{i=n_\mathrm{obs}}^{+ \infty}\frac{(n_\mathrm{bck})^i}{i!}
\mathbf{e^{-n_\mathrm{bck}}}
\sum_{j=n_{\gamma}}^{i}\frac{i!}{j!(i-j)!}p_{\gamma}^j (1-p_{\gamma})^{i-j},
\end{equation}
}
where the first term describes the Poisson probability of observing 
at least \Nobs neutrinos with \Nbck expected background events,
and the second term describes the probability of observing 
at least \Ncoinc coincident gamma-ray flares
{out of the $j\,\ge$\,\Nobs triggers}.
\pgam is the probability to observe a gamma-ray {high state
above a certain threshold ${F_{thr}}$ within a given time window}. 
$P$ defines the post-trial
significance of a set of coincidences observed from one source. 
Trial factors to account for the number of sources 
considered can be easily included using Binomial statistics.
For illustration of Equation~\ref{prob}, let us assume that we observe
\Nobs = 10 neutrinos with a background expectation of \Nbck = 10.
In itself this measurement would not be significant. 
However, if coincidences with $\gamma$-ray high-states are observed
the significance increases as shown in Figure~\ref{significance} for
different $\gamma$-ray probabilities.
So far, limited knowledge is available on \pgam. 
Efforts are on-going to address the issue of estimating an 
upper limit on \pgam for a few interesting sources, 
from a compilation of gamma-ray observations~\cite{tluc} 
and from random or long term monitoring observations 
(e.g. performed by the VERITAS and the MAGIC telescopes). 
We notice that a compilation of existing data is likely biased 
from the availability of measurements triggered by high states 
of emission observed at different wavelengths, which would 
tend to give an overestimation of \pgam and therefore an 
underestimation of the significance of the coincidences.
The probability \pgam is, on average, equal to the average 
high-state rate of an object.
One method for the estimation of the high-state rate is
{based on} the flux frequency distribution of the object,
shown in Figure~\ref{fluxstates} for Mrk\,421. {This distribution
can be interpreted as a stochastic flux-state distribution and can
be well fit by an exponential}. The high-state rate \rhs(${F_{thr}}$)
{above a threshold ${F_{thr}}$
is then given by}
\begin{equation}\label{hsr}
  R_{HS}(F_{thr}) = \frac{
                        \int_{F_{thr}}^\infty e^{b x} dx
                         }
                         {
                        \int_{F_{0}}^\infty e^{b x} dx
                         }
                  = \frac{e^{b F_{thr}}}{e^{b F_{0}}}
\end{equation}
where ${F_{0}}$ is the baseline flux of the object and b is the slope of
the flux distribution.
The relative high-state rate
of Mrk\,421 as derived from this formula is shown 
in Figure ~\ref{hsrate} 
as a function of the chosen threshold $F_{thr}$.
{Due to the bias to high states of the available Mrk\,421 observations,
the high state rate is systematically overestimated here.}
These results will be described in detail in \cite{tluc}.
The estimation for \pgam can be used in Equation~\ref{prob} 
in the case of Blazars, for which $\gamma$-ray data exist
and long-term lightcurves have been compiled.
The expected background rate is the rate of atmospheric neutrinos 
in the sky bins around the selected sources. 
Depending on the source declination and on the choice of the bin size, 
this rate ranges from about 1 to 4 events per year and per source based 
on the AMANDA-II event information and according to the current scheme 
of event reconstruction and selection \cite{achterberg:2006a}.
\vspace{-0.2cm}
\subsection{List of selected sources}
{The first criterion for the selection of sources for the NToO
test run is their variability.
Only sources known or expected to be variable were chosen 
for the test run.
Other criteria are}
the minimal impact on the scientific
plans of 
MAGIC and the possibility to efficiently
organize the independent observation plans.
Target sources are therefore preferably selected among those which
are already included in the scheduled observation program (MAGIC).
Further criteria are their potential for high-energy neutrino emission,
good visibility for MAGIC during the time period of the test run (September--December) and
previous detections at high-energy $\gamma$-rays {or} high probability for
$\gamma$-ray emission.
Sources meeting these requirements are 
Blazars and X-ray binaries.
For these sources the level of correlation between 
high energy neutrinos and gamma-rays 
can be different under different scenarios 
(see for example the cases discussed in~\cite{mq2}). 
%
%
\begin{table}
\begin{center}
\setlength{\tabcolsep}{1.4mm}
\renewcommand{\arraystretch}{0.8}
\begin{tabular}{|llllll|}\hline
		& \rotatebox{90}{\bf\tiny LSI+61\,303}& \rotatebox{90}{\bf\tiny GRS\,1915+105}& \rotatebox{90}{\bf\tiny 1ES\,2344+514}	& \rotatebox{90}{\bf\tiny 1ES\,1959+650} & \rotatebox{90}{\bf\tiny Mrk\,421} \\\hline
\Nbck		& 0.86&1.26 &0.99&0.92&1.51 \\
\Nobs		& 0 & 1 & 1 & 0 & 3 \\
Follow ups 	& 0 & 0 & 1 & 0 & 1 \\
\Ncoinc		& -- & -- & 0 & -- & 0 \\
$F_{thr}$ $[$C.U.$]$& 0.2 & 0.2 & 0.5 & 1.0 & 4.0 \\
\pgam		& -- & -- & -- & $<0.15$ & $<0.05$ \\\hline
\signeu		& 1.0 & 0.7 & 0.6 & 1.0 & 0.2 \\\hline
\end{tabular}
\caption{\label{sourcelist}List of selected sources for the NToO test run. 
Given are preliminary numbers for expected (\Nbck) and observed (\Nobs) neutrino 
triggers, the number of observed coincidences (\Ncoinc), the $\gamma$-ray
high-state probability and the probability \signeu\ for observing \Nobs neutrinos or more.
The error on \Nbck is typically 0.1.
}
\end{center}
\end{table}

\subsection{Results and Interpretation}
During the two months of data taking for the NToO program a total of 5
neutrino event triggers were initiated by AMANDA-II and sent to the MAGIC observatory.
In two cases follow-up observations were performed with the MAGIC
telescope lasting for 1 hour each. For the remaining 3 triggers, the source was not observable
with MAGIC within 24~h following the trigger due to unfavourable
astronomical, moon or weather conditions. 
In Table~\ref{sourcelist} the individual neutrino event counts \Nobs
are given along with
the number of expected neutrino background events \Nbck, the number of
coincident observations with MAGIC, the number \Ncoinc of observed coincident
$\gamma$-ray flares (as defined above) and the $\gamma$-ray flare probability
\pgam derived from Equation~\ref{hsr}.
%
%
The MAGIC follow up observation data has been analyzed with the
standard MAGIC analysis chain \cite{magicanalysis}. 
The sensitivity of MAGIC is sufficient
to detect a $\gamma$-ray flux level of 30\% Crab Units (C.U.) with 5 sigma
significance within 1 hour. It is therefore enough to determine
whether the 2 triggered sources Mrk421 and 1ES2344 were in flaring
state (according to the definition of
flaring state in Table~\ref{sourcelist})
 during the NToO observations.
The analysis yielded an upper limit for 1ES2344 (16\% C.U.) and
a low flux state for Mrk421 (30$\pm$10\% C.U.). No coincident
$\gamma$-ray flaring state has thus been observed.

\section{Discussion and Perspectives}
The NToO strategy was implemented
in a test run involving the AMANDA-II and
the MAGIC telescope for a time period of two-months.
No coincident events have been observed during this test run.
However, the technical feasibility of a NToO strategy
was successfully tested. 
The neutrino trigger information
sent via e-mail has initiated follow-up observations, whenever the sources
were visible and the weather and astronomical (moon/sun) 
conditions allowed the operation of the
MAGIC telescope. 
At the end of the test run, a different communication
infrastructure was also implemented, based on a test client/server
connection, which allows the queuing of follow-up observations using a
similar pipeline as that already used by MAGIC to follow-up GRB alerts.
Perspectively, different event selections will be developed for IceCube.
A search for multiplets with pre-defined significances 
will provide a means for the selection of
flare-like neutrino events.
Furthermore, work is in progress for the analysis
of high-energy neutrino events with the IceCube 22-string 
detector (2007) and with extensions in subsequent years.
These analyzes will possibly be implemented in an IceCube
NToO program in 2008.



\bibliography{litbank_mexico}

\begin{thebibliography}{1}

\bibitem{achterberg:2006a}
A.~{Achterberg~et~al.~for~the~IceCube~collaboration}, 2006.
\newblock arXiv:astro-ph/0611063, accepted for publication in Phys. Rev. D
  (2007).

\bibitem{amanda}
Andr{\'e}s et~al.
\newblock {\em Astropart. Phys.}, 13:1, 2000.

\bibitem{bernardini:2005a}
E.~{Bernardini~for~the~IceCube~collaboration}.
\newblock In {\em Towards a Major network of Atmospheric Cherenkov Telescopes},
  2005.

\bibitem{magicanalysis}
T.~Bretz and R.~M. Wagner, 2003.
\newblock in Proc. of the 28th ICRC, Tsukuba, Japan.

\bibitem{magic}
E.~Lorenz.
\newblock {\em New Astron. Rev.}, 48:339, 2004.

\bibitem{tluc}
M.~Tluczykont et~al.
\newblock In {\em 2nd Workshop on TeV particle Astrophysics, Madison, WI, USA},
  2006.

\bibitem{mq2}
D.F. Torres and F.~Halzen.
\newblock 2006.
\newblock arXiv:astro-ph/0607368, submitted to A\&A.

\end{thebibliography}
\bibliographystyle{plain}

\end{document}